\documentclass{appolb}
\usepackage{epsfig}

\begin{document}
\pagestyle{plain}
\newcount\eLiNe\eLiNe=\inputlineno\advance\eLiNe by -1

\title{Structure and Fine Structure in Multiparticle Production Data at High Energies
}
\author{Wit Busza
\address{Physics Department and Laboratory for Nuclear Science, Massachusetts Institute of Technology, Cambridge, Massachusetts, USA}}
\maketitle
\begin{center}Based on lectures given at the Cracow School of Theoretical Physics, XLIV Course, 2004\\
\end{center}

\begin{abstract}
A summary is given of data on the longitudinal rapidity and pseudorapidity distributions observed in $e^+e^-$, pp, pA and AA collisions at high energies.
The remarkable simplicity and universality observed in the data and its relevance to the study of the high energy density system produced in heavy ion collisions is discussed.
\end{abstract}

There is much data on multiparticle production in high energy collisions.  Reasons for interest in such data are twofold.  First, there is intrinsic interest, particulary for the collision of elementary systems ($e^+e^-$, $p\bar{p}$, pp), on how the initial high energy state evolves into the multiparticle final state.  Second, there is interest in the system that exists between the instant of collision and the production of the final free streaming particles.  Here the key question is whether in high energy collisions of large nuclei an intermediate state is created whose properties are more that of thermalized matter than of non-interacting particles.  If so, is the predicted quark-gluon plasma phase of QCD created and observed?

Today, we know that when two heavy ions collide at ultra-relativistic velocities, for example Au+Au at RHIC energies, the process that takes place is not that of separate nucleon-nulceon collisions immediately producing non-interacting outward streaming particles which give rise to particle spectra that are nothing other than the  superposition of nucleon-nucleon spectra.  There is good reason to believe that in heavy ion collisions at high energies, in a very short time after the initial impact ($\le$ 1fm/c) a very high energy density ($\ge 5GeV/fm^3$), strongly interacting, system is created which then evolves into the observed multiparticle state \cite{Wpa04}.

In these lectures I concentrate on describing the dependence of the longitudinal features of multiparticle production on energy and on the nature of the colliding systems, point out a remarkable universality exhibited by these features over a very broad range of energies and seemingly very different colliding systems, where the intermediate state cannot possibly be the same, and discuss the implications of the observed facts on our understanding of the intermediate state and on its production.

The aim of my talk is to look at the big picture.   Intentionally I am not focusing on details.  Thus for example, the data discussed throughout the talk is for all charged particles without separating them into different species.  Furthermore I am making no attempt to review in an unbiased way all the existing data.  The choice of the data shown is primarily driven by convenience.  I should also mention that this is an expanded version of a talk that I gave at the 20th Winter Workshop on Nuclear Dynamics in Jamaica, 2004.

\begin{figure}
\begin{center}
                 \epsfxsize=7cm
                  \epsfbox{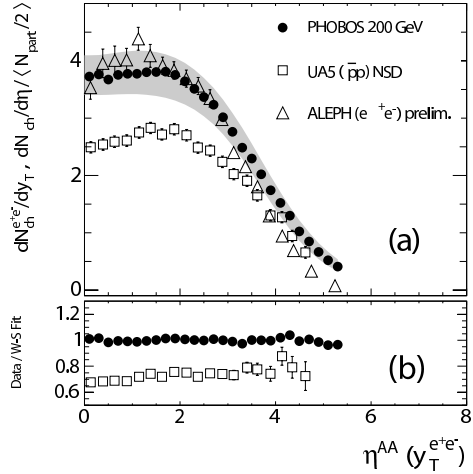}

\caption[]{Pseudorapidity distributions for $e^+e^-$, $p\bar{p}$ and AuAu collisions at 200 GeV\footnotemark[1].  The figure is from \cite{Bac04a}.  In a) the AuAu data is per participant pair\footnotemark[2].  In b) the AuAu and $p\bar{p}$ data are divided by a fit to the $e^+e^-$ data.  The data illustrate the similarity of the shapes for very different colliding systems.}
\label{fig1}
\end{center}
\end{figure}

\begin{figure}
\begin{center}
                 \epsfxsize=12cm
                  \epsfbox{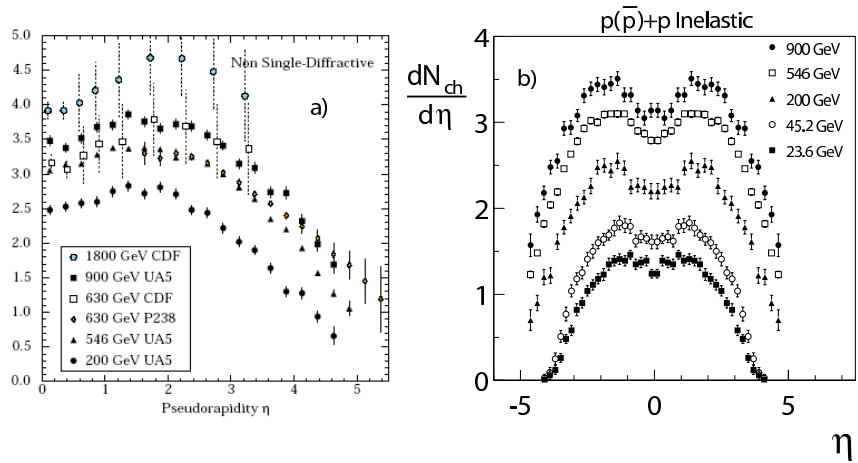}

\caption[]{Energy\footnotemark[1]  dependence of pseudorapidity distributions for pp and $p\bar{p}$ collisions.  a) Distributions for non-single diffractive collisions.  Data taken from \cite{Eid04}.
b) Distributions for all inelastic collisions \cite{Hol04}.  Data taken from \cite{Aln86, Tho77}.  Note: The apparent central plateau and ``double-hump'' structure are not seen in rapidity distributions.  They are a consequence of the Jacobian for transformation between $y$ and $\eta$.}

\label{fig2}
\end{center}
\end{figure}

\begin{figure}
\begin{center}
                 \epsfxsize=6.5cm
                  \epsfbox{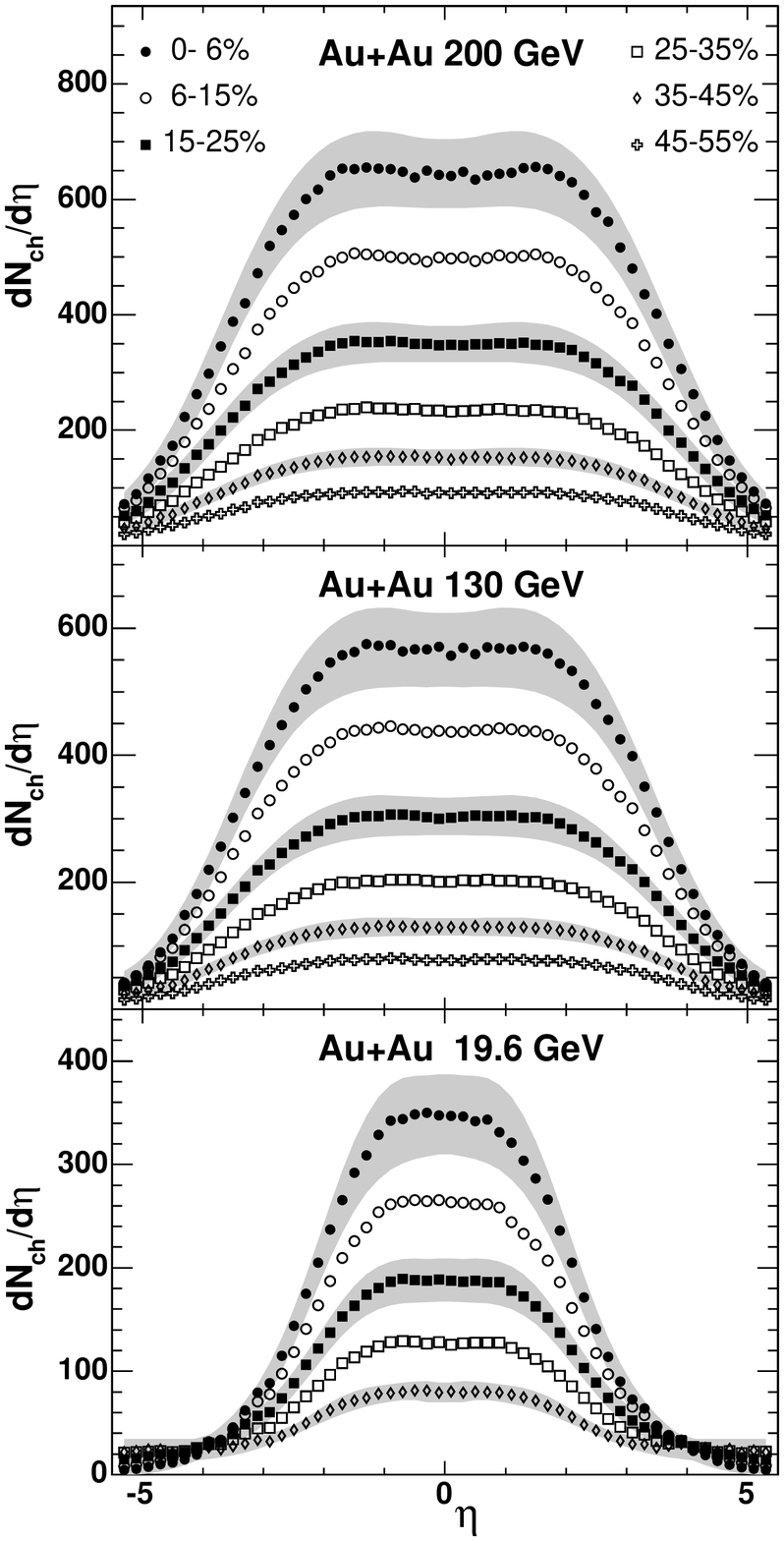}

\caption[]{Energy and centrality (impact parameter) dependence of pseudorapidity distributions for AuAu collisions.  The figure is from the Phobos Experiment at RHIC \cite{Bac04a}, the first systematic study of pseudorapidity distributions in AA collisions over the full $4\pi$ solid angle.  As discussed in the text, the apparent boost-invariant central plateau is misleading.  It is not seen in the true rapidity distributions.  For the 200GeV data the average number of participants\footnotemark[2] corresponding to the various centrality ranges are: 344, 276, 200, 138, 93, 65.}
\label{fig3}
\end{center}
\end{figure}

I start by discussing the overall shape of the observed rapidity distributions in high energy collisions.  Figs 1-3 show pseudorapidity distributions in $e^+e^-$, pp and AA collisions.  The similarity of the shape of the distributions for collisions of such very different systems or which collide with energies\footnote[1]{Unless otherwise stated all energies quoted are the total energy $\sqrt{s}$ in the center of mass system.  For collisions of complex systems such as pA and AA the energy $\sqrt{s_{NN}}$ is normalized to that of a single nucleon in one system colliding with a single nucleon in the other.} that differ by more than an order of magnitude is apparent.  At first sight these distributions and their similarity give the impression that the overall particle production process at high energies is not particularly interesting; that in all cases the particles are simply produced according to the available boost-invariant longitudinal phase space and a limited phase space in the transverse direction.  From this point of view, the pseudorapidity data is misleading.  Although, in general, the pseudorapidity ($\eta = tanh^{-1} \frac{P_l}{P} = tanh^{-1} cos \theta$) is a good approximation of the true rapidity ($y = tanh^{-1} \frac{P_l}{E} = tanh^{-1} \beta$),  the two are not identical.  Differences between the two have a perverse effect on the pseudorapidity distributions, generating a misleading central plateau where there is none in the true rapidity distribution.  Shown for example in figs 4 and 5, the rapidity distributions for collision of symmetric systems such as $e^+e^-$, pp, and AA do not have a significant boost invariant central plateau even at the highest energies studied, and instead have an approximately gaussian shape with a height and width which grow with energy.  This can be seen in fig. 6 where, as an example, true rapidity distributions for AA collisions are shown for a variety of energies.

\begin{figure}[h]
\unitlength10cm
\begin{minipage}[h]{6.5cm}
\begin{center}

\vspace*{-.16cm}
\epsfxsize=6cm

                 \epsfbox{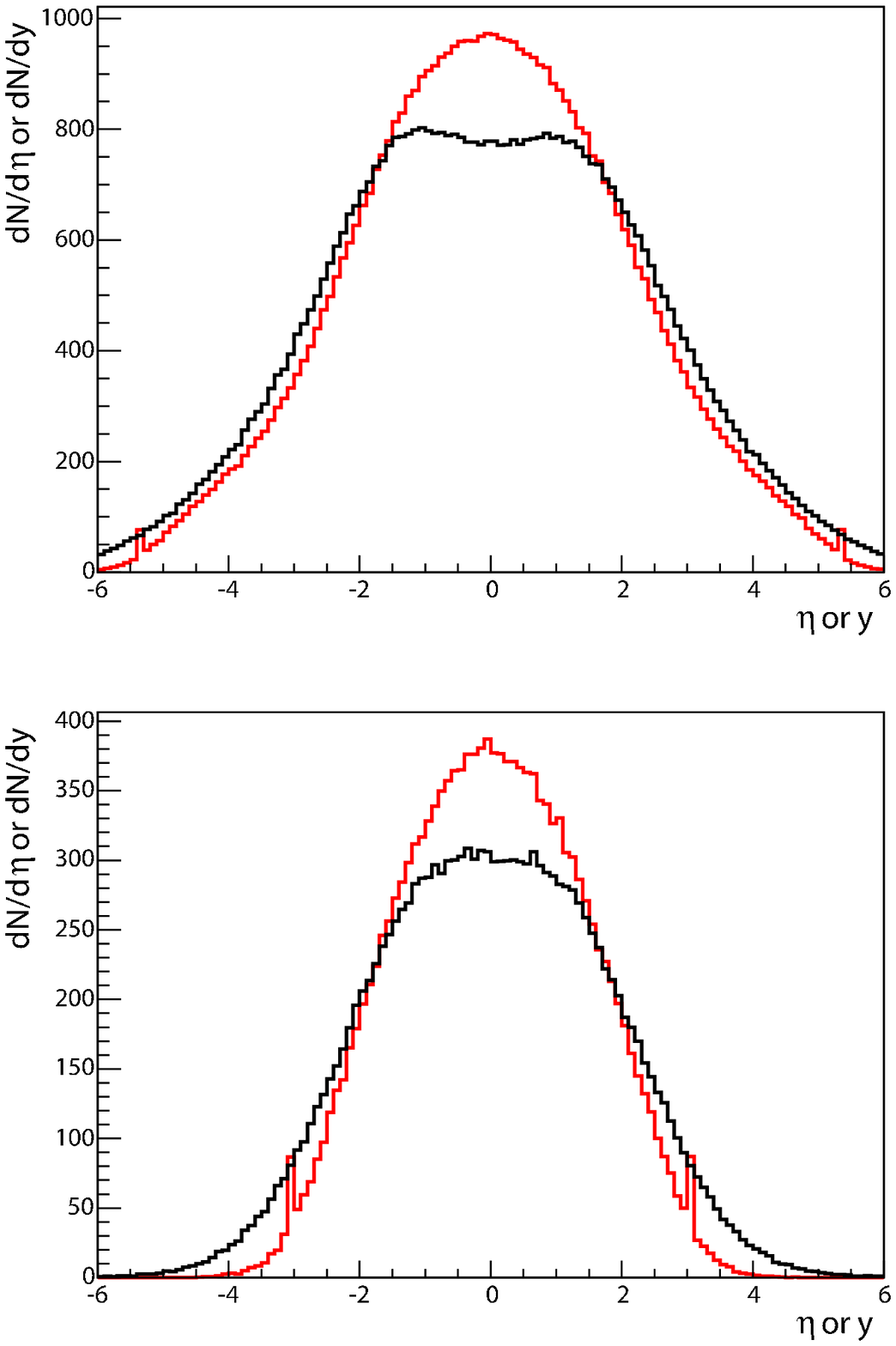}
\vspace*{-.16cm}
\caption[]{Comparison of $dn/d\eta$ and $dn/dy$ for a sample of 19.6GeV and 200GeV simulated events \cite{Woz04} for AuAu collisions using a HIJING Monte Carlo generator.  As can be seen, for a rapidity distribution which has a guassian-like shape the pseudorapidity is almost trapezoidal with an apparent 
boost invariant plateau.}
  \label{fig4}
\end{center}
 \end{minipage}
\hfill
\begin{minipage}[h]{5.5cm}
 \begin{center}

\vspace*{.2cm}
                 \epsfxsize=5cm
                  \epsfbox{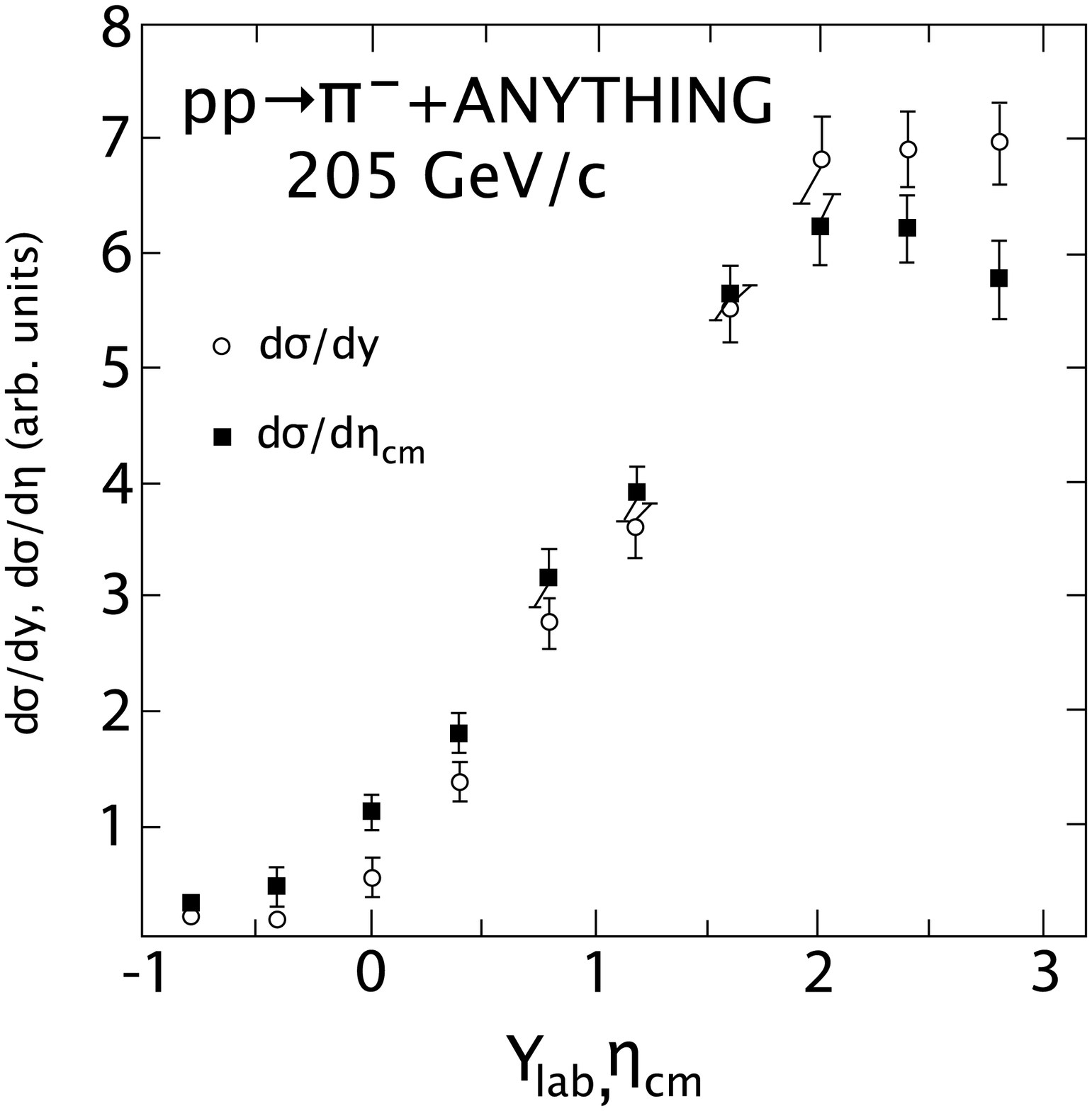}

\vspace*{2.7cm}
\caption[]{Direct comparison of rapidity and pseusorapidity distributions for the same sample of events obtained in a bubble chamber experiment \cite{Whi73}.  The energy quoted is that of the beam with the target at rest.  As expected the pseudorapidity distribution measured in the center of mass system gives a distorted picture of the rapidity distribution, in particular near mid rapidity and the kinematic edges of the distribution.}
\label{fig5}
\end{center}
\end{minipage}

\end{figure}

\begin{figure}
\begin{center}
                 \epsfxsize=9cm
                  \epsfbox{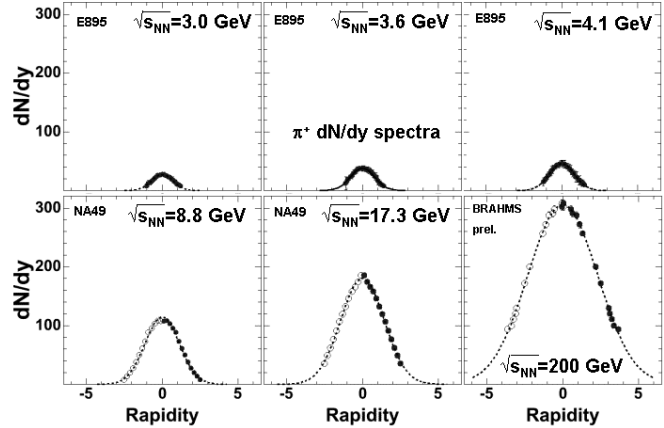}

\caption[]{Examples of rapidity distributions observed in AA collisions at a variety of energies.  The figure is from \cite{Bac04a}.  Note that in contrast to fig 3 $dn/dy$ exhibits no boost invariant central plateau.}
\label{fig6}
\end{center}
\end{figure}

As a first approximation, for both the collision of symmetric and asymmetric systems, the basic distribution is the same; gaussian for the rapidity distribution and trapezoidal for the pseudorapidity distribution.  However in the case of asymmetric colliding systems, such as pA and dA shown in figs 7-9,  the basic distribution is tilted by an amount which depends on the relative number of participants\footnote[2]{In pA and AA collisions the geometry of the collision or impact parameter are characterized either by the centrality of the collision (fraction of total inelastic cross-section, with smaller numbers being more central) or by the number of nucleons participating in the collision.  The latter are calculated using the Glauber model, \ie using the assumption that each nucleon maintains a constant cross-section (the total inelastic nucleon-nucleon cross-section at the incident energy) as it penetrates in a straight line the other nucleus.  The symbol $N_{part}$ or $N^{total}_{part}$ refers to the total number of participants in both colliding systems.  $N_{part}^A$ refers to the number of participants in one of the systems, A in this case.  $N^{total}_{part}$ and ``wounded nucleons'', a concept introduced by Bia{\l}as {\it et al.} \cite{Bia76} are one and the same.  In hadron-nucleus studies rather than using $N^A_{part}$  the symbol $\bar{\nu} = A\sigma_{pp}/\sigma_{pA}$ has been often used.} in the two systems.  The latter fact is most evident by studying fig 10 where for different centralities (\ie for collisions with different impact parameter or with different numbers of participating nucleons) the ratio of the produced particle densities in dAu and pp are plotted as a function of pseudorapidity. The obvious enhancement and suppression of particles seen in fig 10 near the two limits of phase space, I will discuss later.

\begin{figure}
\begin{center}
                 \epsfxsize=10cm
                  \epsfbox{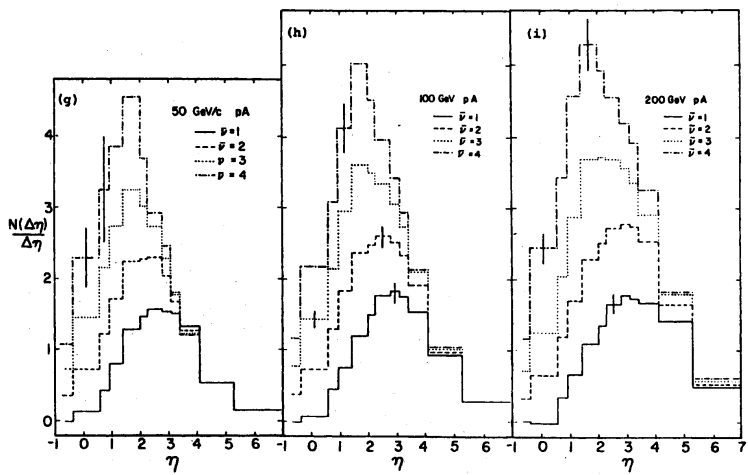}

\caption[]{A-depedence and energy dependence of pseudorapidity distributions in pA collisions.  The data are from Fermilab experiment E178, the first systematic study of pseudorapidity distributions in hadron-nucleus collisions \cite{Bus75, Hal77, Bus77, Eli80}.  $\bar{\nu}$=$N^A_{part}$ is the average number of participants in the nucleus.  The quoted energies are that of the incident proton, with the nucleus at rest.  They correspond to $\sqrt{s_{NN}}$ = 9.8GeV, 13.8GeV and 19.4GeV.}
\label{fig7}
\end{center}
\end{figure}

\begin{figure}
\begin{center}
                 \epsfxsize=7cm
                  \epsfbox{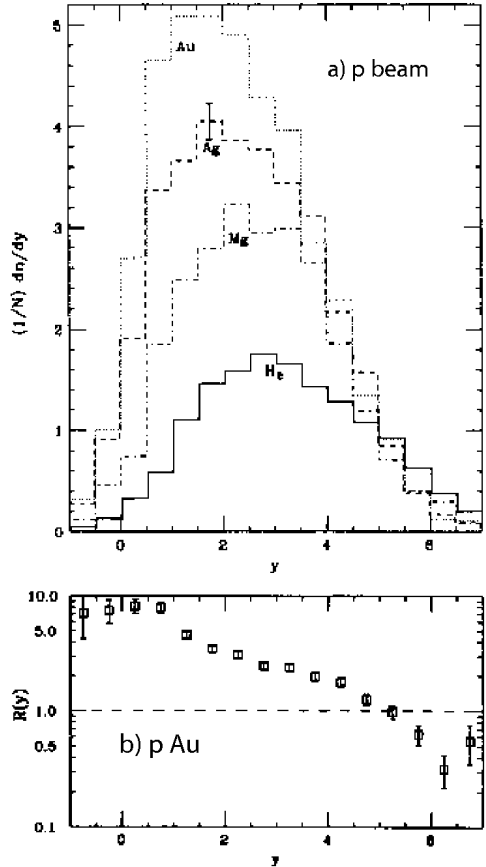}

\caption[]{A-dependence of rapidity distributions for 200GeV protons colliding with various stationary nuclei ($\sqrt{S_{NN}}$ = 19.4GeV) from \cite{Bri90}.  In b.) the ratio $R(y)$ of the particle density in pAu and pp is shown.  Note that $R(y)$ is less than one at the highest rapidity, close to that of the incident proton, and gradually increases as y decreases.  At the lowest rapidities $R(y)$ approaches and even exceeds a value equal to the number of participants in the Au ($\bar{\nu} = N^{Au}_{part} = 3.65$).}
\label{fig8}
\end{center}
\end{figure}

\begin{figure}[h]
\unitlength10cm
\begin{minipage}[h]{6cm}
\begin{center}

\vspace*{-.16cm}
\epsfxsize=5.5cm

                 \epsfbox{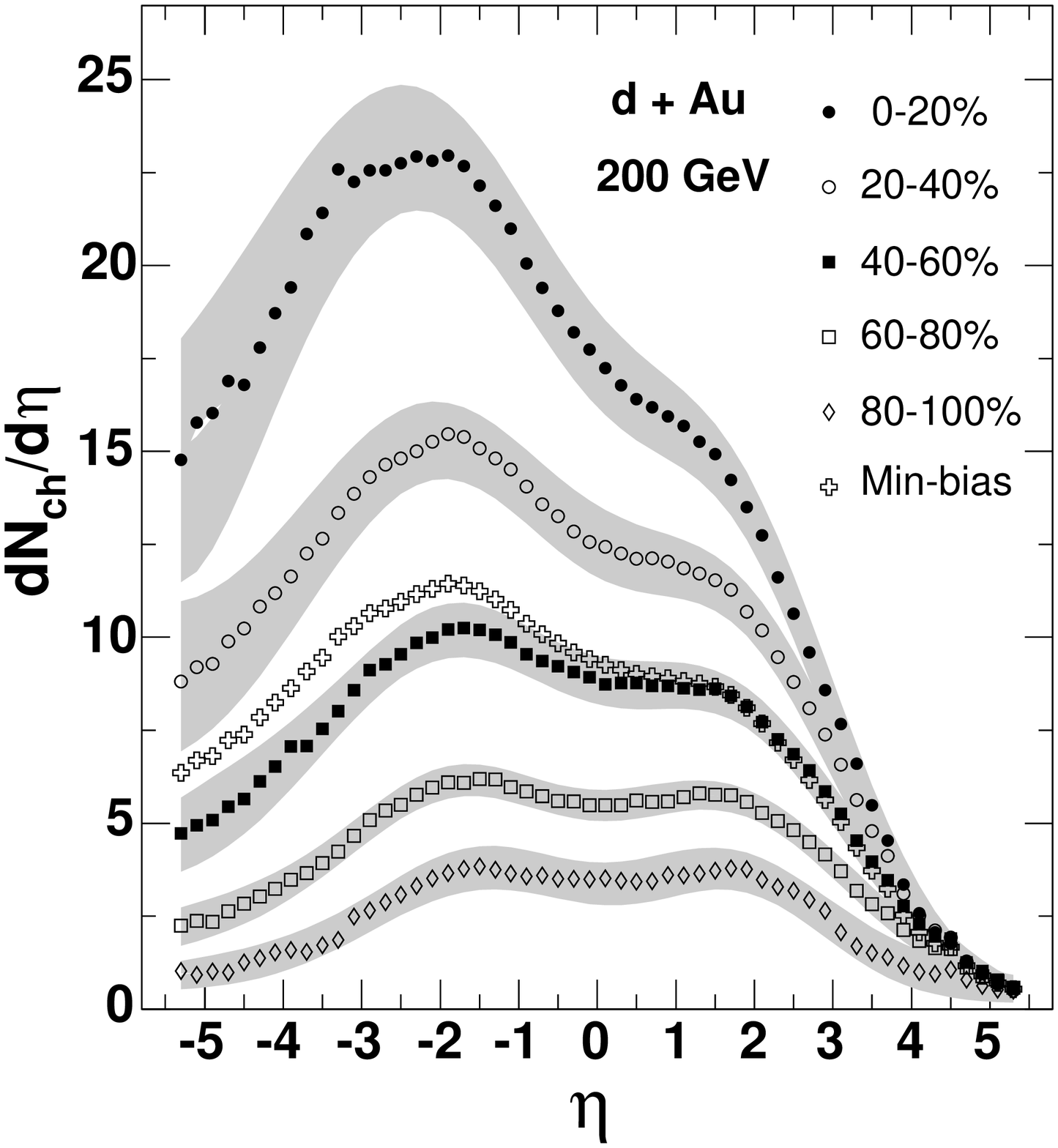}
\vspace*{-.5cm}
\caption[]{Centrality dependence of the pseudorapidity distribution for dAu collisions at $\sqrt{s_{NN}}$ = 200GeV.  The five centrality bins correspond respectively to $N^{Au}_{part}$ = 13.5, 8.9, 5.4, 2.9, and 1.6 and $N^d_{part}$ = 2.0, 1.9, 1.7, 1.4, and 1.1.  The data are from Phobos \cite{Bac04b}.}
  \label{fig9}
\end{center}
 \end{minipage}
\hfill
\begin{minipage}[h]{6cm}
 \begin{center}

\vspace*{-.16cm}
                 \epsfxsize=5.5cm
                  \epsfbox{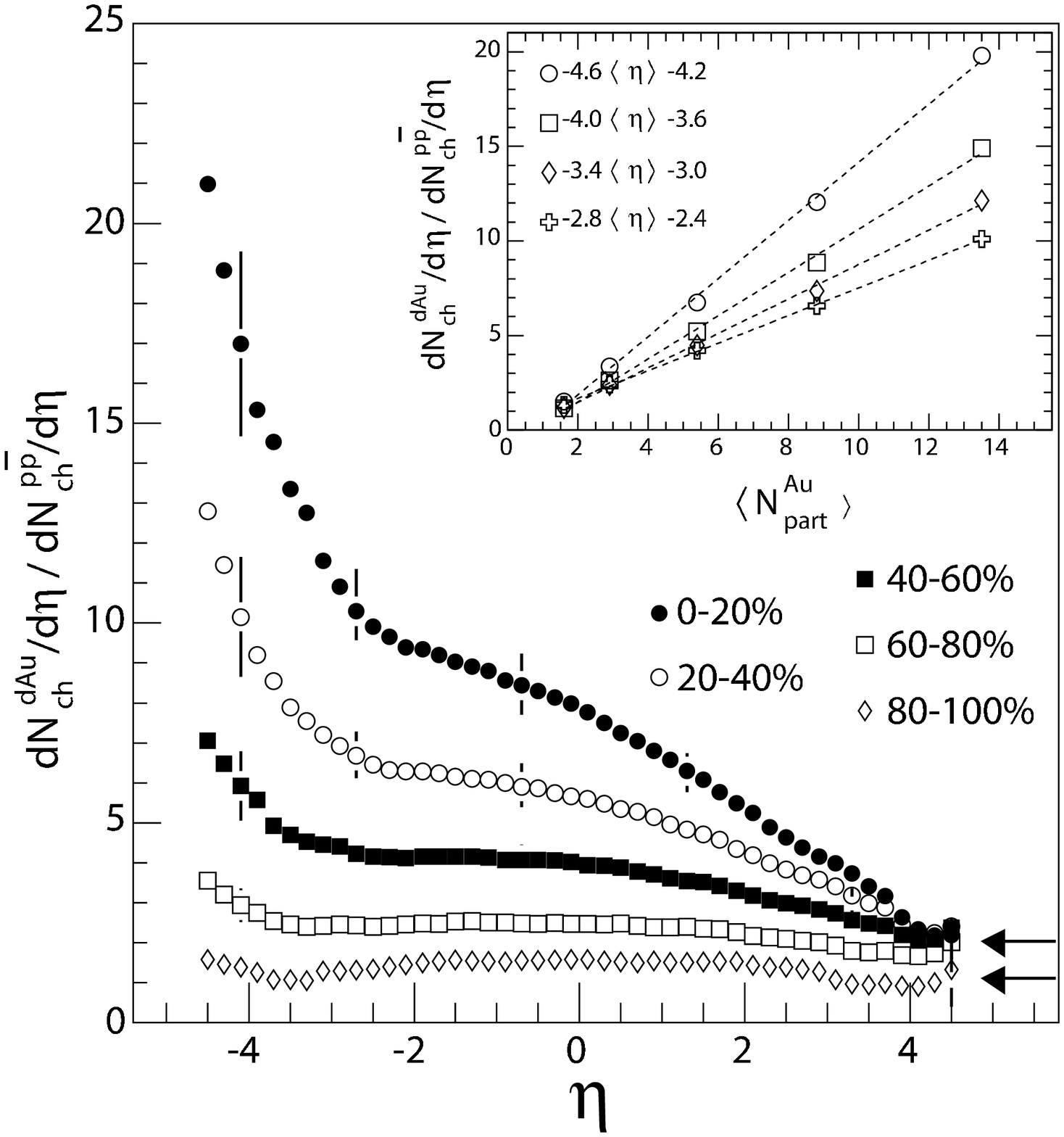}

\vspace*{1cm}
\caption[]{Ratio of particle density produced in dAu and pp collisions as a function of rapidity.  Figure is from Phobos \cite{Bac04b}.  $N^d_{part}$ and $N^{Au}_{part}$ values are listed in figure 9 caption.  The arrows are at a value of $N^d_{part}$ corresponding to the most central and most peripheral collisons.  Note: If one ignores the rise at extreme negative values of $\eta$ the ratio smoothly changes from a value approximately equal to $N^{Au}_{part}$ at one end to $N^d_{part}$ at the other end of the $\eta$ range.}
\label{fig10}
\end{center}
\end{minipage}

\end{figure}

I now turn to the energy dependence of the rapidity distributions (from now on in the discussion of the data I will not differentiate ``rapidity'' and ``pseudorapidity'' unless the difference is of direct relevance).  

The energy dependence of the midrapidity particle density for $e^+e^-$, pp and AA is shown in fig 11.  As can be seen, over the entire range of energies measured to date and for all systems studied, the midrapidity particle density increases as $ln\sqrt{s}$ \cite{Rol04} with no indication of the onset of deviation at the highest energies.

\begin{figure}
\begin{center}
                 \epsfxsize=9cm
                  \epsfbox{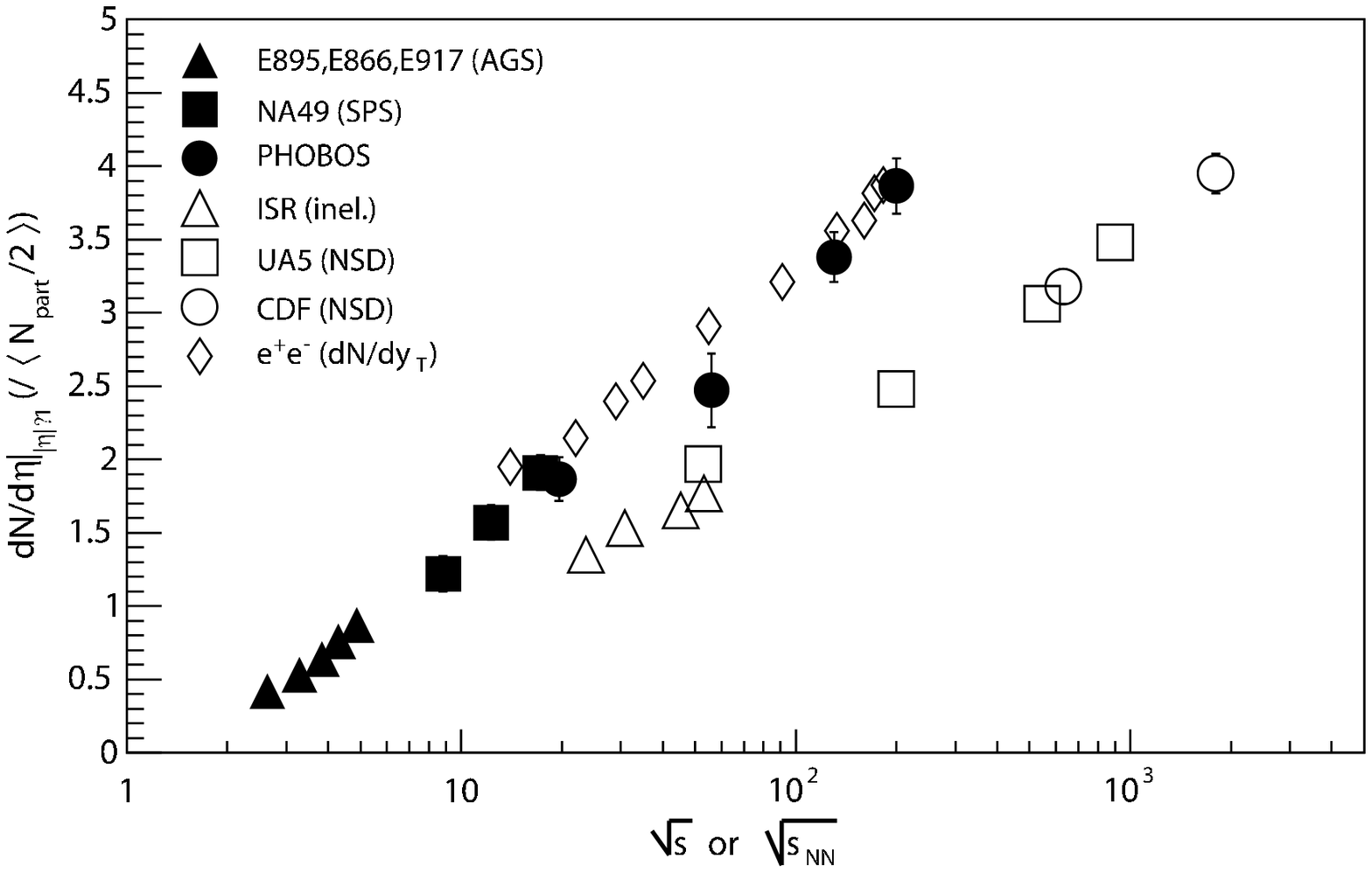}

\caption[]{Data which show that the midrapidity particle density for $e^+e^-$, pp and central AA collisions increases logarithmically with energy.  Figure is from \cite{Bac04a}.}
\label{fig11}
\end{center}
\end{figure}

\begin{figure}
\begin{center}
                 \epsfxsize=11cm
                  \epsfbox{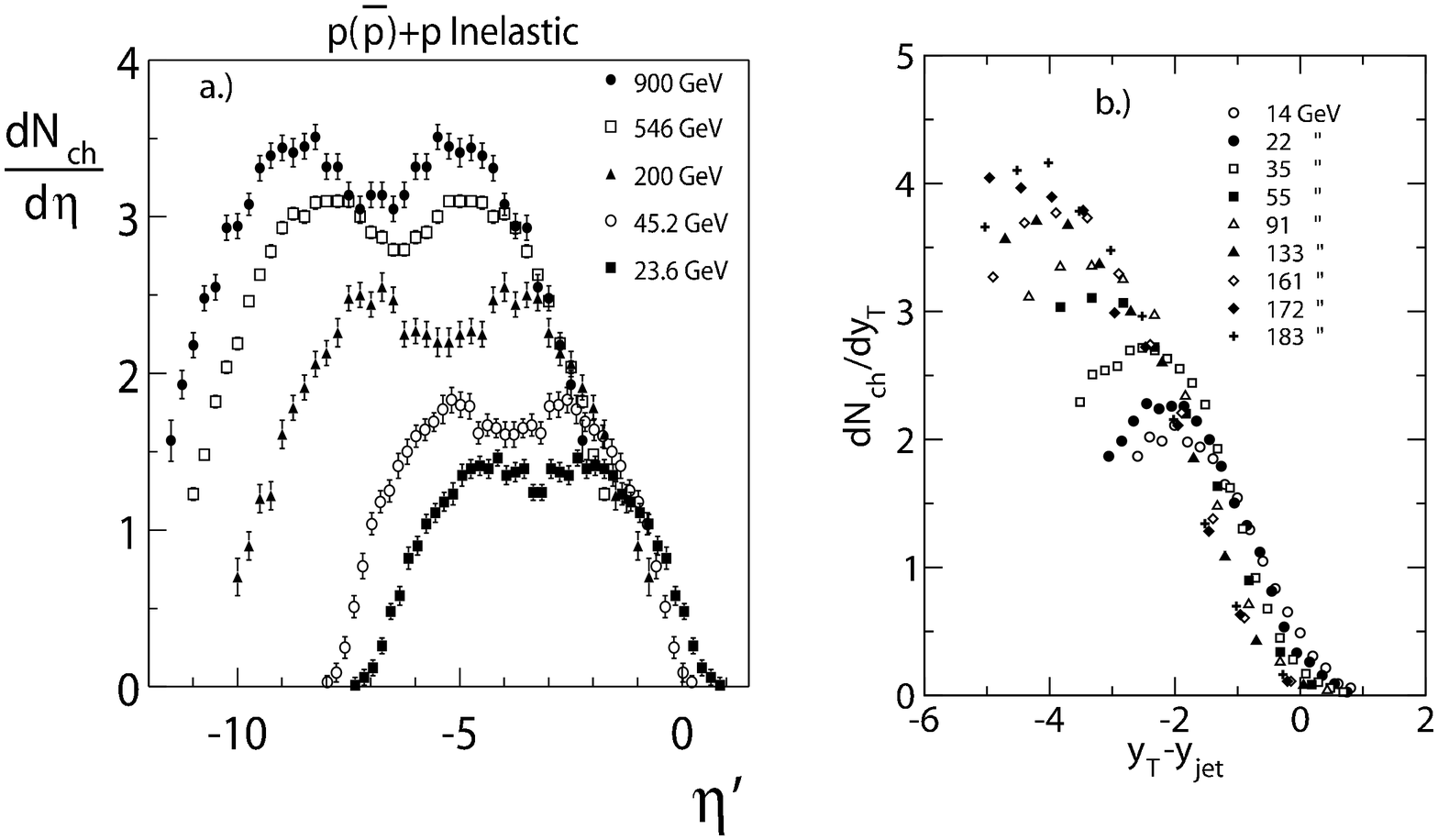}

\caption[]{a.) Pseudorapidity distributions for pp or $p\bar{p}$, (same data as in fig 2b)) plotted in the rest frame of one of the colliding particles and b.) similar distributions for $e^+e^-$ for various energies \cite{Wpa04a}.  The data clearly exhibit the phenomenon of extensive longitudinal scaling discussed in the text.}
\label{fig12}
\end{center}
\end{figure}

The most instructive way of looking at the growth with energy in the longitudinal direction of the rapidity distribution is to compare distributions at different energies in the rest frame of one of the incident particles.  Such comparisons are shown in fig 12 for  $e^+e^-$ and pp, in fig 13 for pA, and in fig 14 for AA collisions.  Clearly in the longitudinal direction the rapidity distribution for all these colliding systems grows with the growth of the available longitudinal phase space, \ie as $ln\sqrt{s}$.  A particularly striking feature of the data is the emergence of a ``limiting curve'' which, as the energy increases, gradually becomes the dominant feature of the pseudorapidity distribution.  This is in contrast to the expectation that at high energies a boost-invariant central plateau would emerge.   The Phobos Collaboration has called this phenomenon extensive longitudinal scaling \cite{Bak04}.  Here I should mention that the best example of extensive longitudinal scaling and the lack of a boost invariant central plateau comes from the Phobos data on elliptic flow in AuAu collisions.  See fig 15.  

The energy dependence of the total number of charged particles produced in the collision of various systems is shown in fig 16.  A $ln^2\sqrt{s}$ dependence for all systems and over the entire range of energies studied is seen.  Again there are no indications of deviation from the trend at the highest energies studies.  From the $ln\sqrt{s}$ dependence of the width and height of the rapidity distribution and the $ln^2\sqrt{s}$ dependence of the total multiplicity, it follows that for given colliding systems the shape of the rapidity distribution does not change with energy.

\begin{figure}
\begin{center}
                 \epsfxsize=9cm
                  \epsfbox{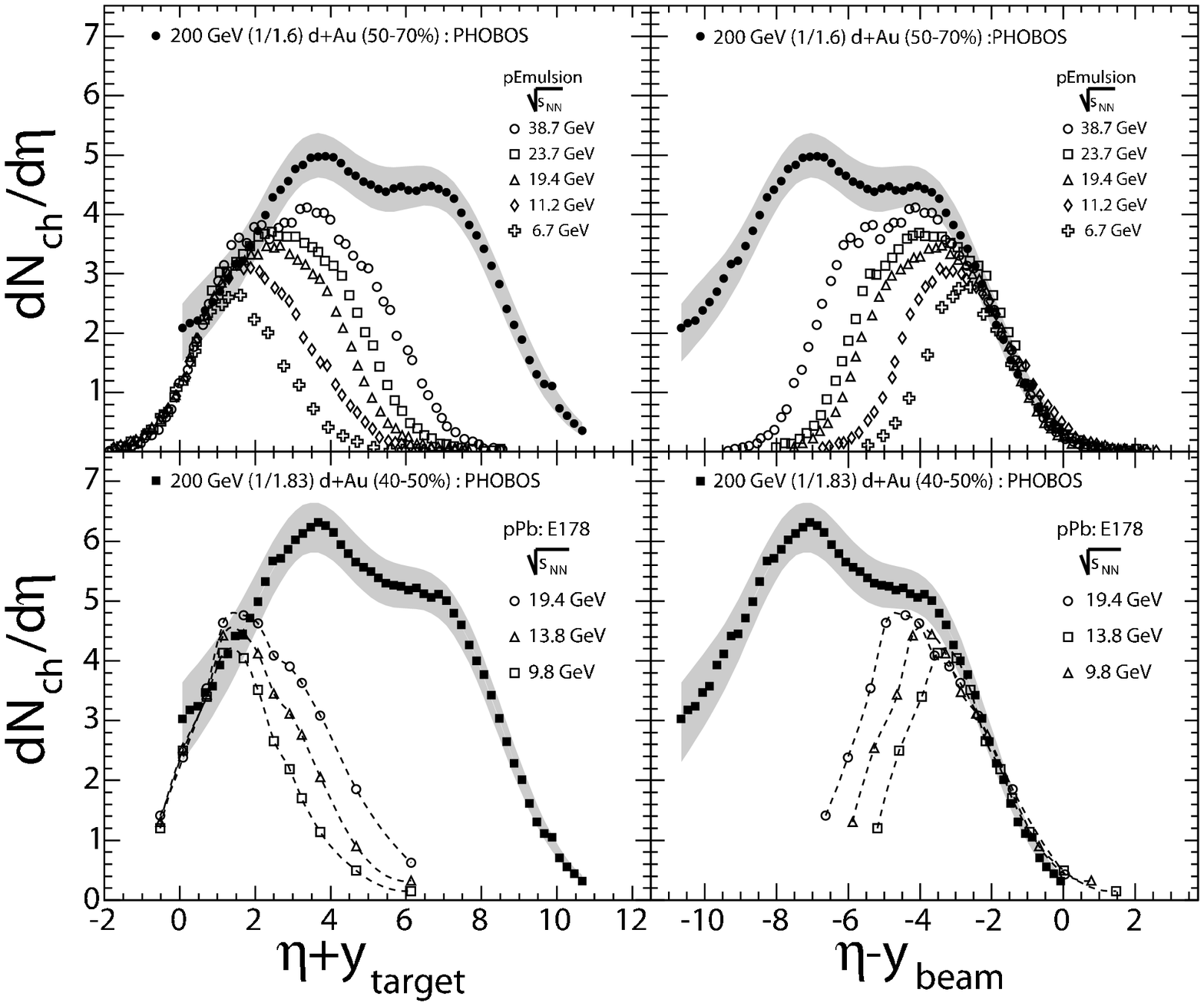}

\caption[]{Examples of extensive longitudinal scaling seen in the rest frame of both the proton and the nucleus in pA collisions.  The figure is from \cite{Bac04a}.}
\label{fig13}
\end{center}
\end{figure}

\begin{figure}
\begin{center}
                 \epsfxsize=9cm
                  \epsfbox{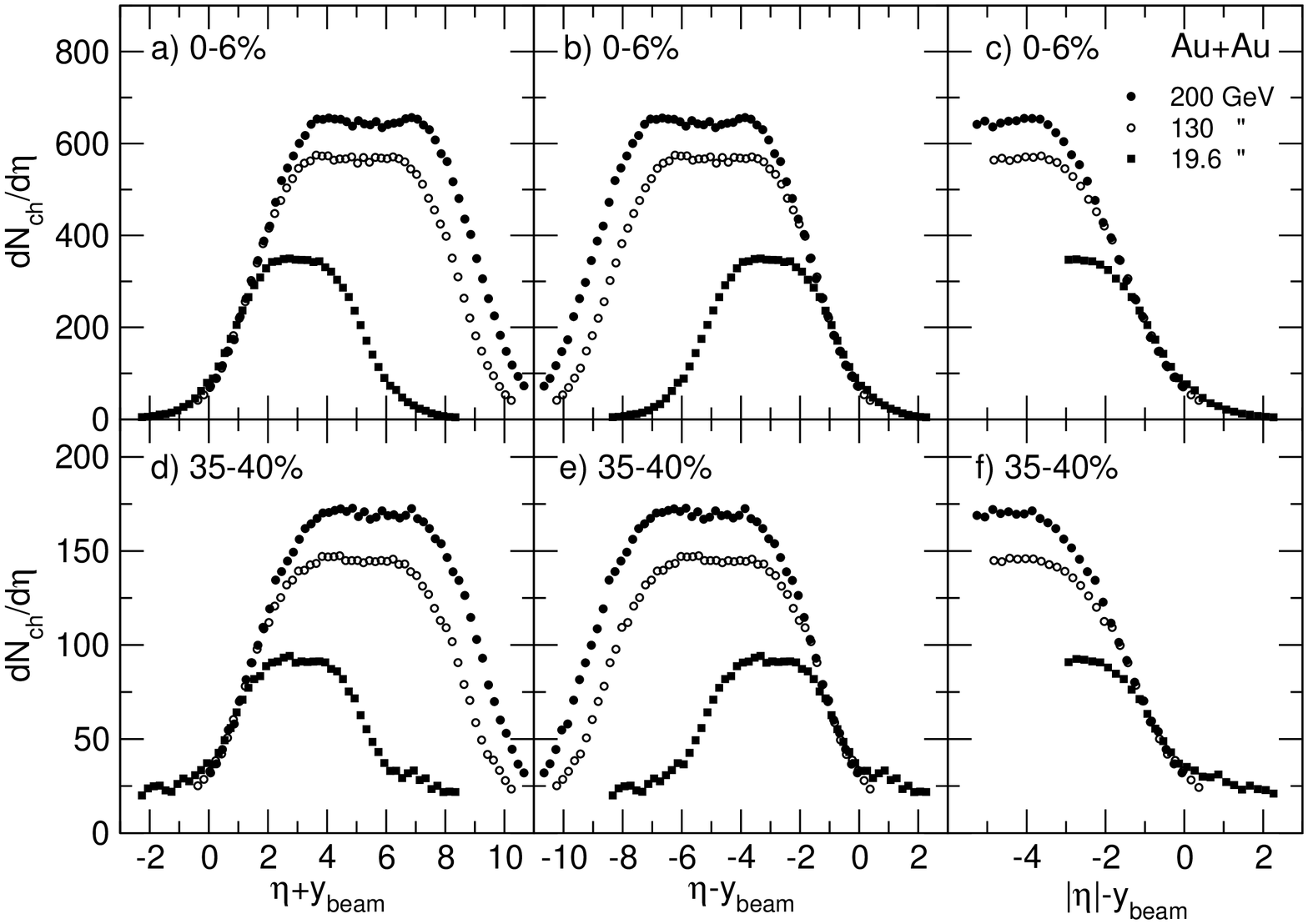}

\caption[]{Examples of extensive longitudinal scaling seen in the rest frame of both nuclei in AuAu collisions.  The figure is from Phobos, \cite{Bac04a}.}
\label{fig14}
\end{center}
\end{figure}

Next we turn to the dependence of the rapidity distribution on the nature of the colliding systems.  Not surprisingly the rapidity distribution does depend on the colliding systems.  This is evident if one compares, for example, the rapidity distributions for AuAu collisions at different impact parameter as shown in fig 3, or $e^+e^-$ and pp data in fig 1 with dAu data in fig 9.  From the earliest studies of hadron-nucleus collisions (see for example fig 17) through recent studies of AA collisions at RHIC energies we know that in such collisions the key parameters that determine the shape and integral of the rapidity distribution are the energy of the collision, as already discussed, and the number of participants (wounded nucleons in the language of Bia\l as {\it{et al}} \cite{Bia76}.) in each of the colliding systems.  As a first approximation the influence of each quantity is independent of the other quatities.  As can be seen in the data, the participants in both colliding systems influence the rapidity distribution in a very specific way which is most evident if we look at pA and dA data, for example figs 9 and 10.  Clearly each participant in both the gold nucleus and in the deuteron has an effect on the produced particle density which extends over the entire rapidity range.  Furthermore, the data suggest that, as a first approximation, the contribution of each participant to the total particle density linearly decreases with the rapidity gap between the participant and the produced particles\footnote[3]{At the time of submission for publication of these lecture notes I recieved the draft of a paper ``Wounded nucleon model and Deuteron-Gold Collisions at RHIC'' by A. Bia{\l}as and W. Czy{\.{z}}.  I refer the reader to this paper.  The authors make similar observations as I do in this part of my talk, however they develop the ideas much further than I do.}.

\begin{figure}
\begin{center}
                 \epsfxsize=11cm
                  \epsfbox{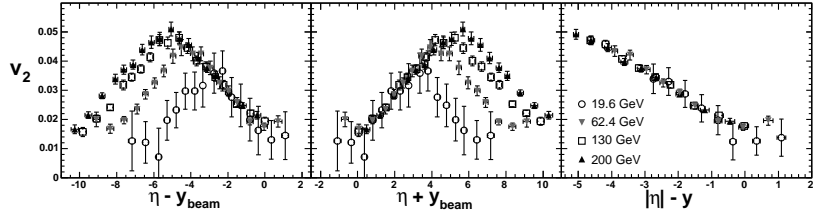}

\caption[]{A spectacular example of extensive longitudinal scaling seen in the Phobos data on elliptic flow in AuAu collisions.  The figure is from \cite{Bac04a}.}
\label{fig15}
\end{center}
\end{figure}

\begin{figure}
\begin{center}
                 \epsfxsize=9cm
                  \epsfbox{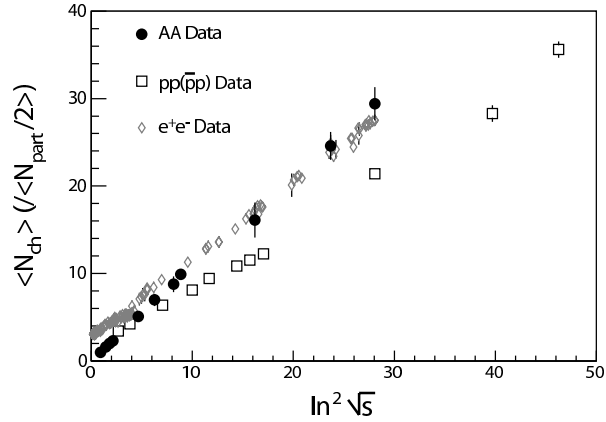}

\caption[]{A compilation of data \cite{HenSte}, which show that the total charged particle multiplicity in $e^+e^-$, pp, $p\bar{p}$ and AA collisions all scale with energy as $ln^2\sqrt{s}$.}
\label{fig16}
\end{center}
\end{figure}

As a consequence of this dependence of the rapidity distribution on the number of participants in each system and consistent with the data, the total charged particle multiplicity is proportional to the total number of participants.  As can be seen in fig 18 for pA and dA this $N^{total}_{part}$ scaling is well represented by $N_{ch} = 1/2 N^{total}_{part} N_{pp}$
where $N_{ch}$ and $N_{pp}$ are the total charged particle multiplicity in pA and pp collisions at the same energy.

\begin{figure}
\begin{center}
                 \epsfxsize=9cm
                  \epsfbox{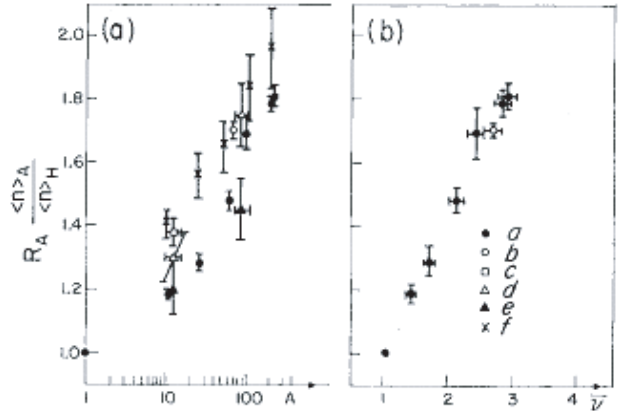}

\caption[]{The ratio of the total number of charged particles produced in hadron-nucleus collisions to that in hadron-proton collisions at the same energy plotted (a) as a function of A and (b) as function of $\bar{\nu}$.  The figure is from \cite{Bus75}.  The fact that this ratio, for a variety of systems and energy of collision, was found to be only a function of $\bar{\nu}$ showed for the first time that the number of participants in the collision is a key parameter in the description of high energy collisions involving nuclei.}
\label{fig17}
\end{center}
\end{figure}

\begin{figure}
\begin{center}
                 \epsfxsize=9cm
                  \epsfbox{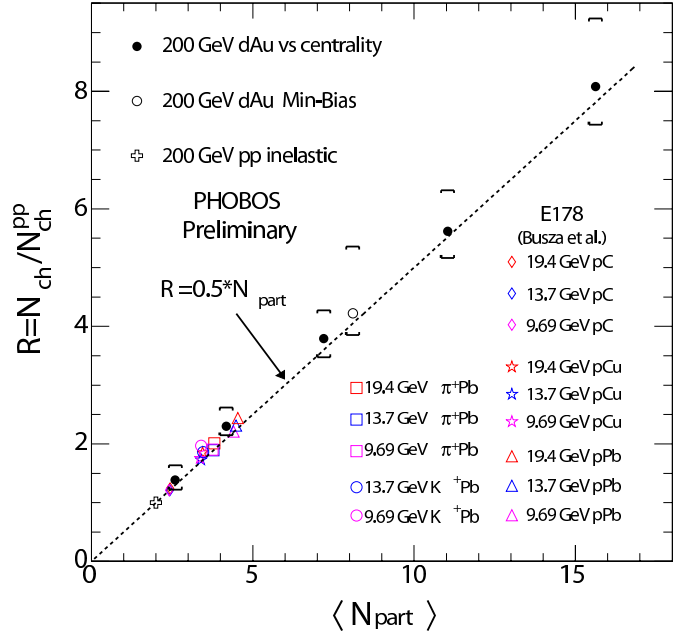}

\caption[]{The ratio of the total number of charged particles produced in hadron-nucleus and deuteron-nucleus collisions to that in hadron-proton collisions at the same energy, plotted as a function of the total number of participants in the collision.  The data are from the Phobos \cite{Bac04a} and E178 \cite{Bus75, Hal77, Bus77, Eli80} experiments.  The data clearly exhibit universal participant scaling.}
\label{fig18}
\end{center}
\end{figure}

In AA collisions $N^{total}_{part}$ scaling is also observed, as is evident from fig 19, however the constant of proportionality is slightly higher.  On closer examination of figs 8 and 10 it is obvious that the linear dependence of the particle density at a given rapidity on the number of participants and on the rapidity gap between the produced and incident particles is only a crude first approximation.  There are clearly two other effects present.  In the fragmentation region of the nucleus there is an enhancement of particles, and in the fragmentation region of the proton there is a depletion.  The latter fact has been thoroughly studied in pA collisions and some results are shown in figs 20 and 21.

\begin{figure}
\begin{center}
                 \epsfxsize=9cm
                  \epsfbox{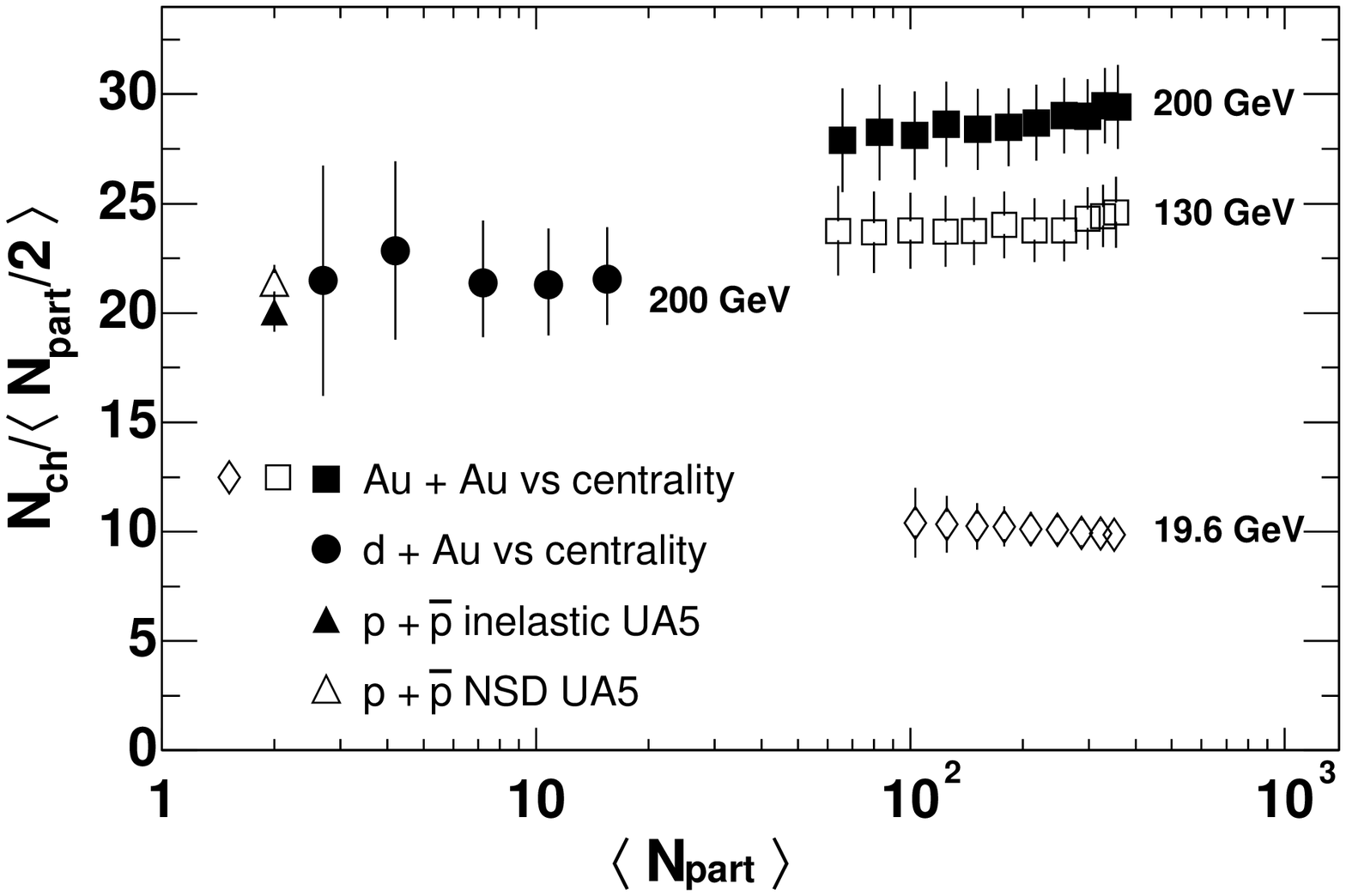}

\caption[]{The difference in $N_{part}$ scaling observed in AuAu and in dAu or pp collisions.  The figure is from \cite{Bac04a}.  The most likely cause of the difference is a difference in the energy available for particle production.  See text.}
\label{fig19}
\end{center}
\end{figure}

\begin{figure}
\begin{center}
                 \epsfxsize=9cm
                  \epsfbox{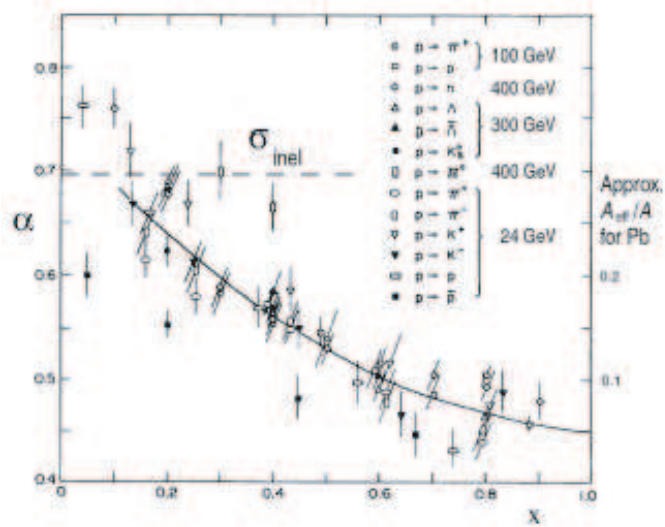}

\caption[]{A-dependence of the production of particles in the forward two units of rapidity for the process pA$\rightarrow$hX (from \cite{Bar83}). $\alpha$ is the power of A in a $\sigma_o A^{\alpha}$ parameterization of the invariant cross-section.  $x\equiv p_l/p_{incident}$.  Note: The fact that $\alpha$ is independent of energy reflects a consistency with the hypothesis of limiting fragmentation as well as with extensive longitudinal scaling discussed in the text.  For the fastest particles, $\alpha$ is seen to be reduced by almost one third from its value for the total inelastic cross-section.  This suggests that the nucleus is completely absorbing from the point of view of the fastest produced particles.}
\label{fig20}
\end{center}
\end{figure}

\begin{figure}
\begin{center}
                 \epsfxsize=9cm
                  \epsfbox{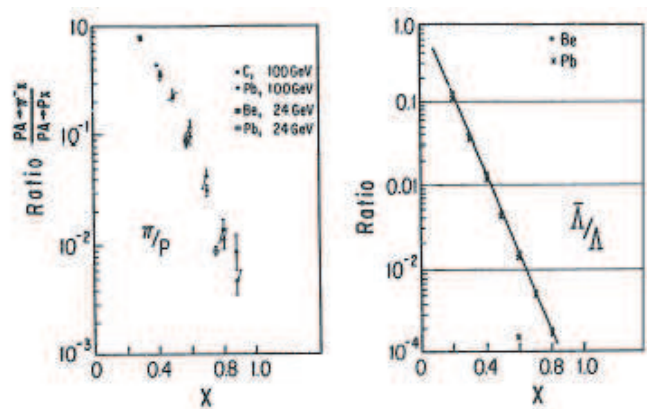}

\caption[]{pA data from \cite{Bus91} showing that the ratios of particles produced in the forward two units of rapidity do not depend on A.  $x\equiv p_l/p_{incident}$.  Note that this data, as that in fig 20, suggests that fast forward particles are only produced on the periphery of the nucleus.  The center of the nucleus appears to be completely absorbing for such particles.}
\label{fig21}
\end{center}
\end{figure}

\begin{figure}
\begin{center}
                 \epsfxsize=9cm
                  \epsfbox{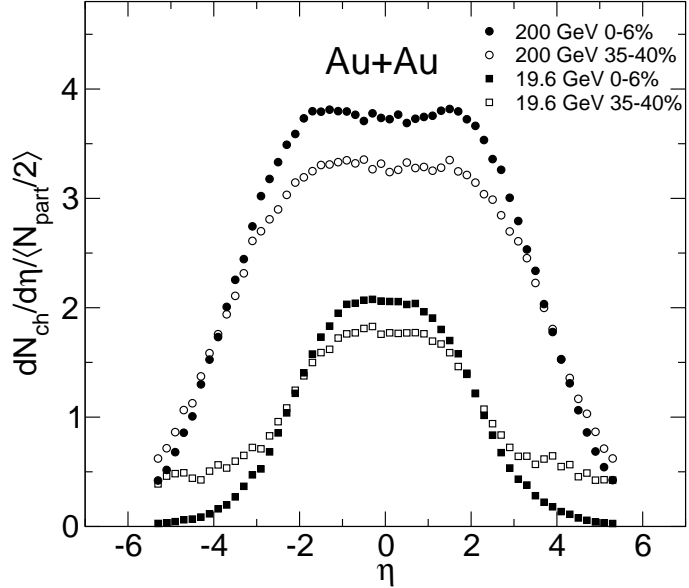}

\caption[]{Comparison of pseudorapidity distributions for central and peripheral AuAu collisions at the same energy.  Note that although the total multiplicity per participant is the same for different impact parameters the shapes of the distributions are centrality dependent.  The figure is from Phobos \cite{Bac04a}.}
\label{fig22}
\end{center}
\end{figure}

Finally in fig 22 we show a direct comparison of the pseudorapidity distributions for AuAu collisions at two different impact parameters, normalized to the same number of total participants, and in fig 23 the centrality dependence of the midrapidity density of AuAu collisions at two energies.  We see that, as discussed earlier, the total number of particles produced per participant is independent of centrality but that the detailed shape does depend on it .  Furthermore, we see that it depends on it in a very specific, energy independent way.  As I will discuss later, the observed centrality dependence in AA collisions is consistent with that in pA and dA.

Having shown the key features of the data I am now in a position to discuss them.  I should stress here that the discussion that follows reflects how I see the facts.  I am making no attempt to balance my view with those of others, nor am I discussing phenomenological models proposed to explain various aspects of the data.  They are beyond the scope of these lectures.

If we look at the ``big picture'', and are not confused by details, in the multiparticle production data for such varied systems as $e^+e^-$, pp and AA at relatively low and high energies, we see an amazing simplicity and universality.  I am tempted to conclude that this universal ``structure'' seen in the distribution of the produced particles in longitudinal phase space reflects some common underlying physics.  Superimposed on the universal structure there is ``fine structure'' which becomes apparent at the next level of focus.  The latter is a consequence of some aspects of the mechanism which are specific to particular systems in collision.  To be more specific, I now return to the data and point out the ``structure'' and ``fine structure'' that I see in it.

\begin{figure}[h]
\unitlength10cm
\begin{minipage}[h]{6cm}
\begin{center}

\vspace*{-.16cm}
\epsfxsize=6cm

                 \epsfbox{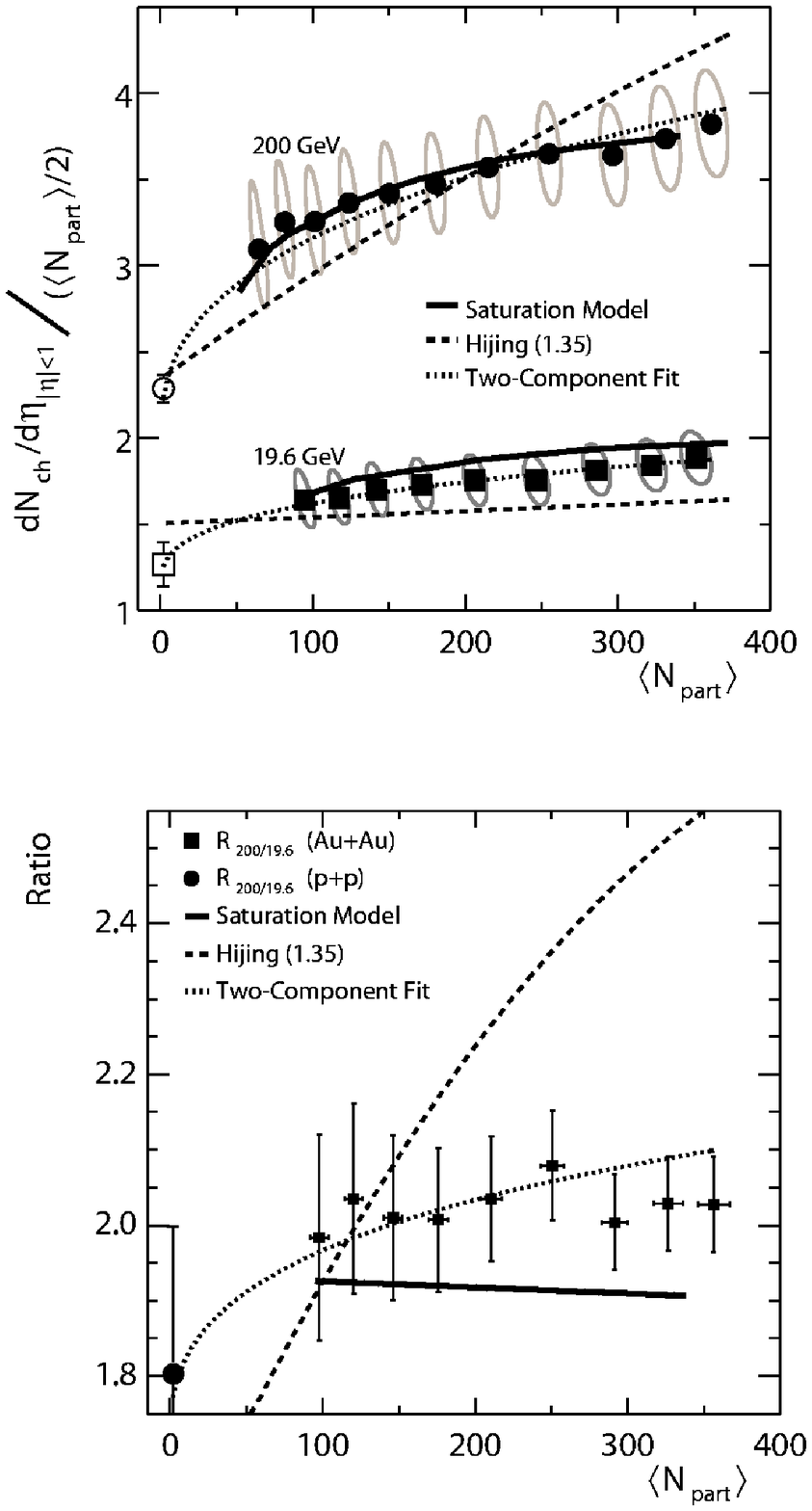}
\vspace*{.5cm}
\caption[]{Figures from \cite{Bac04a} which show that the centrality dependence of the midrapidity density of particle production in AuAu collisions is independent of the energy of the collision.  It is yet another example of the universal behavior of multiparticle production; the separate and independent role played by the energy and the colliding systems.}
  \label{fig23}
\end{center}
 \end{minipage}
\hfill
\begin{minipage}[h]{6cm}
 \begin{center}

\vspace*{-.16cm}
                 \epsfxsize=6cm
                  \epsfbox{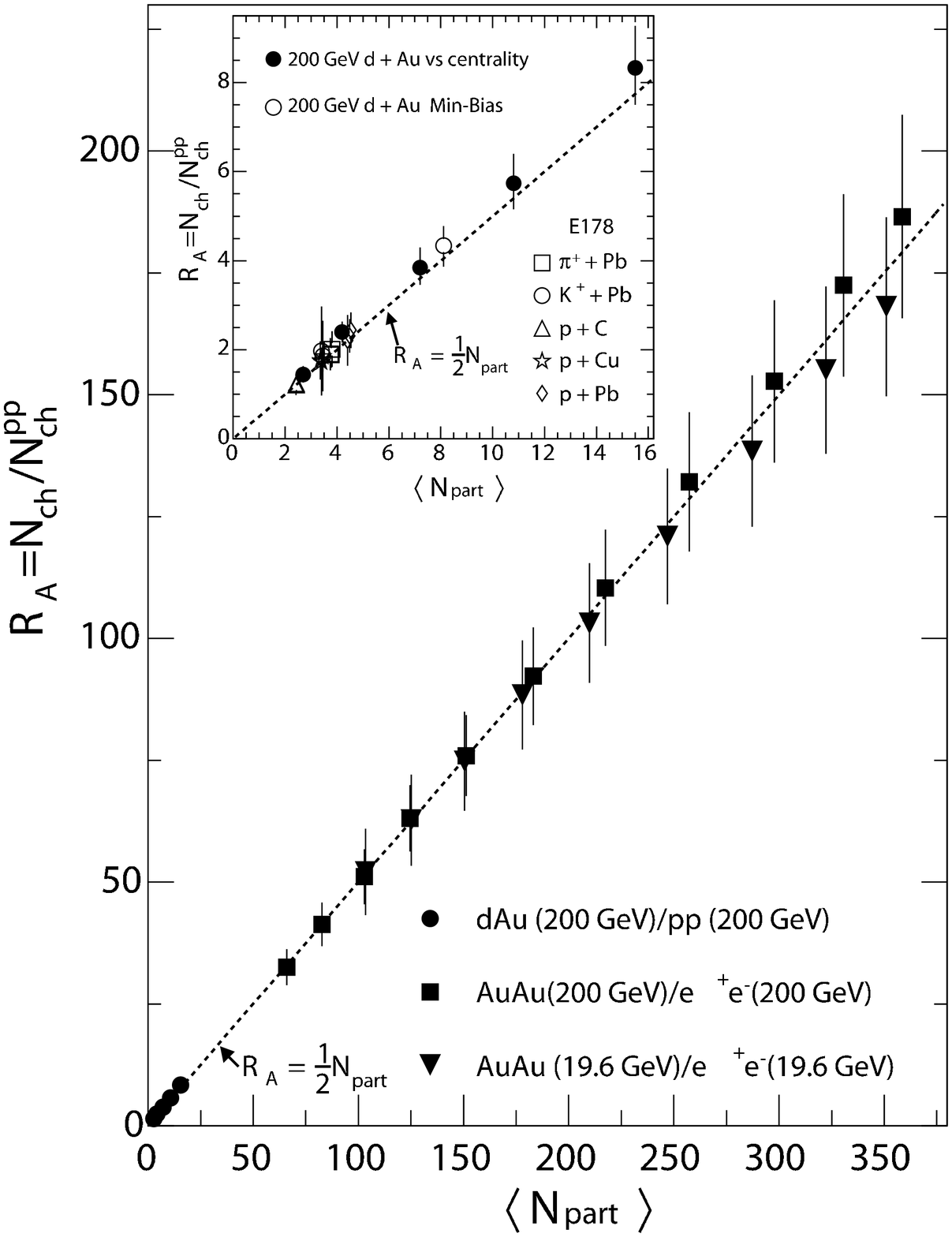}

\vspace*{1.6cm}
\caption[]{Universal $N_{part}$ scaling.  Prediction of the dependence of the total number of charged particles as a function of the total number of participants for hadron-nucleus and nucleus-nucleus collisions, based on the universal ``structure'' observed in multiparticle production, and discussed in the text.  Note that the data covers a range of $N_{part}$ from 2 to 350 and energy from 10GeV to 200GeV.  It is a compilation of Phobos \cite{Bac04a} and E178 \cite{Bus75, Hal77, Bus77, Eli80} data.\cite{Nou04}}
\label{fig24}
\end{center}
\end{minipage}

\end{figure}

The ``structure'' or common features that I see in multiparticle production at lowest magnification are as follows.

1.) The longitudinal rapidity distributions depend seperately on the energy and the nature of the colliding systems. \ie the distributions factorize.  For given colliding systems the shape of the distribution is approximately the same for all energies.

2.) For all processes at a given available energy the rapidity distribution is the same basic distribution adjusted for the number of participants in the two colliding systems.  Probably the best representation of this basic distribution is that observed in $e^+e^-$ collisions (and therefore probably also that in $q\bar{q}$ collisions).
For AA collisions, where most of the energy is available for particle production, the basic distribution is the $e^+e^-$ distribution at the same energy ($\sqrt{s}=\sqrt{s_{NN}}$).
In pp or $p\bar{p}$ collisions only about half the energy goes into particle production, the rest goes into a leading baryon, and thus the basic distribution is approximately that in $e^+e^-$ for $\sqrt{s}=1/2\sqrt{s_{NN}}$.  The latter also applies for pA and dA since in these collisions most of the participants collide with only one nucleon.

3.) Each participant in the collision contributes to the overall rapidity distribution a particle density which is proportional to the basic distribution, with a constant of proportionality which decreases linearly, from one to zero with increasing rapidity gap between the participant and the produced particles.  This naturally leads to universal $N^{total}_{part}$ scaling.

4.) From 2.) and 3.) and the observed shape of the distributions for the collision of symmetric systems, it follows that the basic rapidity distribution is gaussian-like (trapezoid-like pseudorapidity distribution).

5.) From 1.) and the fact that the width of the rapidity distribution increases with energy as longitudinal phase space \ie as $ln\sqrt{s}$, it follows that a.) the mid-rapidity density increases with energy as $ln\sqrt{s}$, b.) the total multiplicity increases as $ln^2\sqrt{s}$, and c.) extensive longitudinal scaling is satisfied.

At higher magnification we begin to see other features in the data which I consider to be the ``fine structure'' superimposed on top of the ``structure''.  Some examples are as follows.

a.) In pA and AA collisions there is an enhancement of particles with rapidity close to that of the incident nucleus.  The enhancement is most likely a consequence of either some intranuclear cascading or of Fermi motion, or both.

b.) In pA and AA collisions there is a suppression of the fastest particles, those produced with rapidity close to that of the proton.  [As a parenthesis I wish to point out that this suppression may prove rather interesting.  The fact observed in fig 20 that the A-dependence of the most forward produced particles drops by almost $A^{1/3}$ and in fig 21 that the $\Lambda$ and $\bar{\Lambda}$ production have the same A-dependence suggests that the production of these particles occurs only on the periphery of the nucleus.  \ie that the center of the nucleus is almost completely absorbing.  Although in a completely different $p_t$ domain, is it possible that this suppression is related to the jet-quenching seen in AA collisions?]

c.) Per participant, the rapidity distributions for central and peripheral AA collisions have the same area but not the same shape (see fig 22).  Futhermore the change of shape is energy independent (see fig 23).  The most likely explanation for the change of shape is that it is a consequence of the interplay of three effects: a.) and b.) discussed above, and most significantly the fact that, for the same number of total participants, the distribution of the participants in the two colliding nuclei is not the same for central and peripheral collisions (\ie central collisions are made up of collisions of symmetric systems whilst peripheral collisions of asymmetric systems).

d.) There are differences in the rapidity distributions of different species or for selected particles with high transverse momentum.  These I consider beyond the scope of this talk.

From the fact that we are able to summarize multiparticle production in a simple, yet quantative, way it follows that we can make predictions.  One prediction is that for pp, pA, dA, and AA at all energies, the normalized total multiplicity $R=\frac{N_{ch}}{N_{basic}}$ is proportional to the total number of participants, with a constant of proportionality of 0.5, where for pp, pA and dA, $N_{basic}$ is the multiplicty in pp collisions at the same energy.  For AA, $N_{basic}$ is the multiplicity in $e^+e^-$ at the same energy.  This prediction can be tested with existing data.  It is shown in fig 24.  Considering that the data spans an energy range from 10GeV to 200GeV and $N_{part}$ from 2 to 350, and includes $\pi$A and KA data, the agreement is spectacular.

Another prediction we can make is that in the upcoming CuCu run at $\sqrt{s_{NN}} =200 GeV$ at RHIC for the same number of participants the pseudorapidity distribution will have the same total integral as in AuAu at $\sqrt{s_{NN}}=200 GeV$ but it will have a slightly higher particle density at midrapidity.  This follows from the fact that for the same number of participants the collisions in CuCu will have a more symmetric distribution of participants in the two colliding nuclei.

\begin{figure}
\begin{center}
                 \epsfxsize=8cm
                  \epsfbox{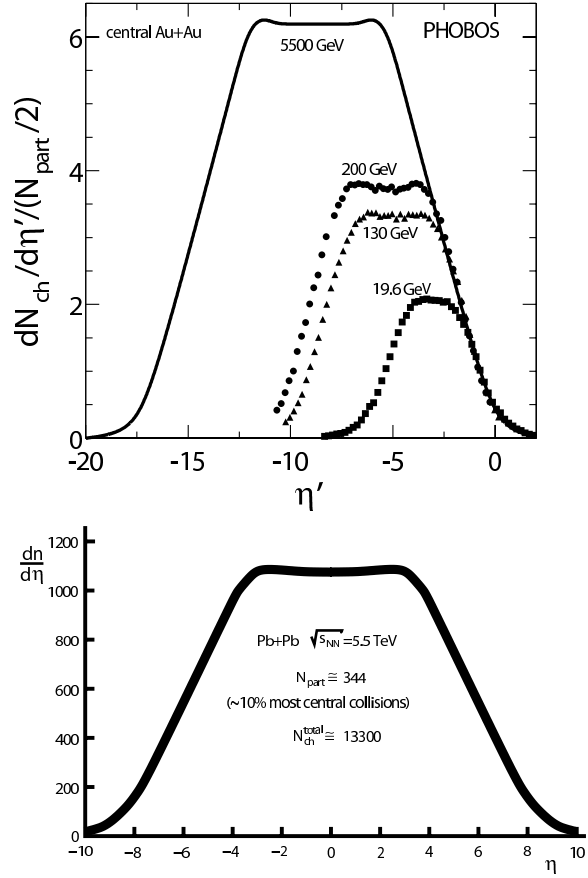}

\caption[]{Prediction of the pseduorapidity distribution for the production of all charged primary particles that will be seen at LHC for the more central PbPb collisions \cite{Ver04}.  In a) the predicted distributions are shown together with lower energy data in the rest frame of one of the nuclei.  In b) it is shown in the center of mass frame.  The prediction is based on extrapolating to LHC energies Phobos data on the mid-rapidity density, total number of produced particles, and on extensive longitudinal scaling.  In short it is based on the ``structure'' seen in multiparticle production data at lower energies.}
\label{fig25}
\end{center}
\end{figure}

Finally, in fig 25, we predict the pseudorapidity distribution and the total number of charged particles that will be seen at LHC.  Should this latter prediction prove wrong, in my opinion, it will signal the onset of some fundamentally new process occurring in AA collisions at LHC energies.

\begin{figure}
\begin{center}
                 \epsfxsize=8cm
                  \epsfbox{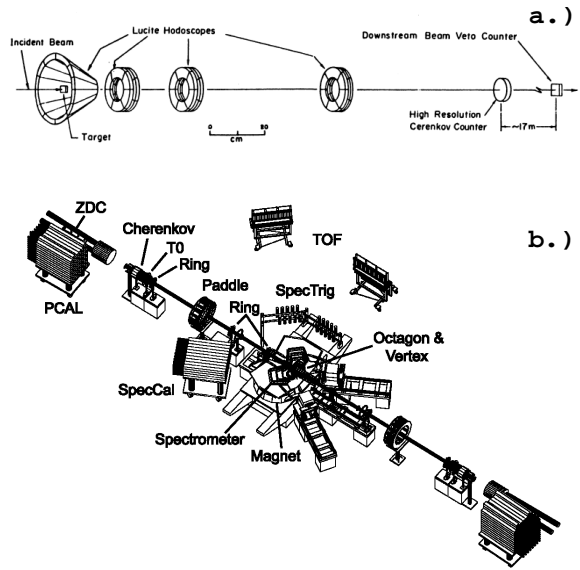}

\caption[]{Diagrams of the two experiments from which most of the data shown in this talk have been obtained.  a.) Fermilab experiment E178  b.) Phobos experiment at RHIC.}
\label{fig26}
\end{center}
\end{figure}

This brings me back to an issue I raised at the beginning of my talk.  There is little doubt that the processes that take place in multiparticle production in such different collisions as $e^+e^-$, pp and AA over such a broad range of energy must be fundamentally varied and the produced intermediate states must  also be very different.  How is it then that the final outcome does not reflect this variety?  I see two possibilities.  The first is that somehow the longitudinal rapidity distribution of all the produced particles is totally insensitive to the details of the process of multiparticle production.  In my opinion this is unlikely  since the rapidity distributions, though universal and relatively simple, are not trivial.  They do not, for example, simply reflect phase space.  The second possibility is that the rapidity distributions are determined locally in the very early phases of the collision process, basically by the structure of the incoming states, and that subsequent processess such as the evolution of any intermediate state formed and final hadronization neither significantly influence the number of particles produced nor their distribution in rapidity.  These are different possiblities with profound consequences on the interpretation of the phenomenology of multiparticle production, and so must be understood.

I have tried to show in these lectures the beautiful simplicity and universality exhibited by the rapidity ditributions in multiparticle production at high energies.  Clearly nature is trying to help us understand what actually happens during these seemingly complicated processes.  I have no doubt that the final correct theoretical description of AA collisions at ultrarelativistic velocities will automatically contain or predict the structure and fine structure seen in multiparticle production data that I have attempted to describe in this talk.

\section{Appendix}
A large fraction of the data shown in my talk is from two rather similar experiments separated in time by 30 years.  The first is Fermilab experiment E178 \cite{Bus75, Hal77, Bus77, Eli80}.  It was a fixed target experiment which studied $\pi$, K, and p-A multiparticle production with 50, 100 and 200GeV/c momentum beams.  In essence it was a hodoscope of \u{C}erenkov detectors covering almost the complete solid angle around the target.  The second is the Phobos experiment at RHIC \cite{Bac04a}.  It is a collider (symmetric in energy) experiment which to date has studied multiparticle production for AuAu collisions at $\sqrt{s_{NN}}$=19.6, 55.9, 62.4, 130.4, and 200GeV, and pp and dAu collisions at 200GeV.  In essence it is a hodoscope of silicon detectors covering almost the complete solid angle around the collision point.
I thought it would be of interest at the end of my talk to show disgrams of these two experiments.  I do so in fig 26.
\section*{Acknowledgments}
I thank my colleagues from experiment E178 at Fermilab and from Phobos at RHIC for numerous discussions related to the subject of this talk, and above all for the measurement of a large fraction of the data presented in this talk.  I thank Miklos Gyulassy for valuable comments related to fig 10.  I also wish to thank the organizers of the school in Zakopane for their hospitality and in particular Andrzej Bia\l as for stimulating discussions.


\begin{thebibliography}{99}
\bibitem{Wpa04a} References to this compilation can be found in \cite{Bac04a}.
\bibitem{Bak04} The name ``extensive longitudinal scaling'' was first suggested by Mark Baker.
\bibitem{HenSte} References to the data plotted in this compilation can be found in \cite{Bac04a}.  I thank Conor Henderson for producing this figure and Peter Steinberg for bringing to my attention the fact that the total multiplicity in AA and $e^+e^-$ collisions at the same energy are almost identical.
\end{thebibliography}
\end{document}